\documentclass[prb,twocolumn,showpacs,groupedaddress,letterpaper]{revtex4-1} 

\usepackage{amsmath}  
\usepackage{amsfonts} 
\usepackage{graphicx} 
\usepackage{xcolor}

\begin{document}

\title{Recommendations for the use of notebooks in upper-division physics lab courses}

\author{Jacob T. Stanley$^1$ and H. J. Lewandowski$^{1,2}$}
\affiliation{$^1$Department of Physics, University of Colorado, Boulder, CO 80309, USA}

\affiliation{$^2$JILA, National Institute of Standards and Technology and University of Colorado, Boulder, CO, 80309, USA}

\begin{abstract}
The use of lab notebooks for scientific documentation is a ubiquitous part of physics research. However, it is common for undergraduate physics laboratory courses not to emphasize the development of documentation skills, despite the fact that such courses are some of the earliest opportunities for students to start engaging in this practice. One potential impediment to the inclusion of explicit documentation training is that it may be unclear to instructors which features of authentic documentation practice are efficacious to teach and how to incorporate these features into the lab class environment. In this work, we outline some of the salient features of authentic documentation, informed by interviews with physics researchers, and provide recommendations for how these can be incorporated into the lab curriculum. We do not focus on structural details or templates for notebooks. Instead, we address holistic considerations for the purpose of scientific documentation that can guide students to develop their own documentation style. Taking into consideration all the aspects that can help improve students' documentation, it is also important to consider the design of the lab activities themselves. Students should have experience with implementing these authentic features of documentation during lab activities in order for them to find practice with documentation beneficial. 
\end{abstract}

\maketitle
\section{Introduction}

Scientific documentation is a vital component in the process of research. These records, documented in physics laboratories all around the world, constitute the foundational information for essentially all the published experimental results found in physics journals. Without thoughtful and thorough records of experimental progress, it is difficult to imagine physicists successfully navigating the complexities of today's research frontier. Given its importance, we should actively be teaching scientific documentation skills to physics students.

To further motivate the importance of teaching documentation, several recent reports have emphasized the need for educational improvements that focus on engaging undergraduate students in skills that are similar to those of professional scientists.\cite{AmericanAssociationofPhysicsTeachers2014,PCAST2012,NSF1996} The use of lab notebooks to perform authentic scientific documentation is identified as one of these skills.

Effective development of students' documentation skills requires understanding what constitutes authentic documentation and how to incorporate it into a lab course. There have been research efforts, outside of physics, that address the incorporation of scientific writing into lab courses, but much of this work has focused on summative forms of writing, such as lab reports.\cite{Moskovitz2011,Jackson2006,Lerner2007,Alaimo2009,Reynolds2012,Ruzycki2015} Most of the work that directly addresses the development of notebook documentation skills can be found in chemistry\cite{Bowers1926,Gould1927,Mortensen1927,Kanare1985,Pendley1997} and engineering\cite{Hughson1979,McCormack1991} education. Some of this literature places strong emphasis on the role of documentation for patent support and as a result is formulaic.\cite{Hughson1979,Kanare1985,Renfrew1989}

Although some of the recommendations from this body of work could prove useful in the context of physics, it is not clear which aspects actually translate. In our view as educators, experimental physicists, and physics education researchers, the physics lab environment has distinctly different concerns and considerations that those of a chemistry, biology, or engineering lab.\cite{Hoskinson2014} For this reason, we thought it prudent to focus on the documentation practices used in physics to inform how to address the use of notebooks in physics lab courses.

In the physics education literature, there has been little work done that addresses documentation. A few exceptions include case studies of historical notebooks\cite{McKnight1967} and descriptions of lab courses that emphasize the importance of notebook documentation, but do not provide details about practice or lab class implementation.\cite{Dayton1962,Miller1965a,Graetzer1972} However, there have been some instructor efforts to design activities that engage students in understanding and developing documentation skills.\cite{Atkins2013,Salter2013} Our goal for this work is to add to these efforts.

In our previous work, we interviewed physics graduate student researchers to understand the educational experiences researchers have with lab notebooks that help them develop their documentation skills. We found that despite the apparent importance and ubiquity of scientific documentation, many had not developed a practice of scientific documentation that was suitable for their research until after they had started in their graduate research groups. Most of them described their undergraduate lab courses as being poor preparation for maintaining these lab records---generally, scientific documentation was not taught in these courses.\cite{Stanley2016a} There are a number of reasons why this could be the case. For example, developing documentation skills may not be a high priority for physics lab courses (in particular, introductory lab courses) and thus it is not established as a learning goal. Alternatively, if developing documentation skills is in fact a learning goal, the course may not be explicitly incorporating documentation-based activities into the curriculum. Regardless, we believe lab courses should take an active approach in motivating and incorporating the use of lab notebooks so that students start developing effective documentation skills earlier. This will not only be helpful for students who go on to do graduate research but also those that pursue other professional avenues that involve technical documentation but are not necessarily centered on research.

The purpose of this paper is to provide physics instructors with an outline of some of the main features of documentation in experimental physics research and recommendations for incorporating the use of notebooks into a lab course that are based on this authentic documentation. Our recommendations are based on (1) our own extensive experience with performing documentation in physics research, (2) our own experience in teaching lab courses, and (3) interviews with graduate student researchers that focused on how they perform scientific documentation in their work and how they developed these documentation skills over the course of their education. The recommendations are the the result of a qualitative synthesis of researchers' interview responses, which we have evaluated and justified using our own expertise. 

In these interviews, we have addressed a range of aspects of the researchers' approach to documentation: the purpose notebooks serve in their research, the format of their notebooks, the degree of detail of their entries, the types of information that they record, the frequency with which they write in and reference their entries, limitations and difficulties that they confront, and other things to consider when recording information.

Some aspects of the researchers' documentation were idiosyncratic and highly variable, but others proved to be quite consistent. Understanding both universal and variable features of authentic scientific documentation can be helpful to inform what should and should not be the focus of instruction. The recommendations we present are for the consideration of instructors when deciding: (1) how to frame the role of notebooks in their course, (2) how to align the use of notebooks with the goals of the lab activities, and (3) how to evaluate the quality of the students' notebook entries and provide feedback.

The remainder of the paper consists of the following: overview of the interview process in Sec. \ref{sec:inter}; a synthesis of the interviewees' stated purpose for documentation in Sec \ref{sec:purpose}; a synthesis of the general characteristics of the interviewees documentation practice in Sec. \ref{sec:principles}; recommendations to instructors that include framing, lab activity design, and evaluation/grading in Sec. \ref{sec:recommendations}; and finally a brief summary in Sec. \ref{sec:sum}.

\section{Interviews}
\label{sec:inter}

The interviewees were 13 physics graduate students (six women, seven men) at the University of Colorado Boulder. They represented a range of years, with the most junior being in their second year and the most senior being in their sixth. All interviewees had spent at least six months doing research in their graduate program---10 of which had spent at least three years doing research in their group. More information about the participants can be found in our previous work.\cite{Stanley2016a}

All interviewees were actively involved in research in one of the ``table-top'' experimental physics subfields (i.e., those that involve daily, hands-on experimental activities with in-house equipment). The breakdown was seven in atomic, molecular, and optical physics; three in condensed matter physics; two in biophysics; and one in plasma physics. 

We chose to focus on these subfields because the daily activities and hands-on nature of the research are most similar to the activities of students in upper-division lab courses, therefore the interviewees' use of notebooks was likely to be the most relevant to the use of notebooks in a lab course.

These interviews were part of a broader effort to explore the scientific documentation of physics graduate student researchers. Each interview consisted of three different parts: (1) lab context, in which we probed the nature of the experiment and work dynamic the interviewee experienced; (2) features of their documentation, in which we probed the specifics of what, how, and why the interviewee recorded the information they did; and (3) educational background, in which we probed the interviewee's experience with scientific documentation at various periods in their education. We have previously used these interview data in a research capacity to examine the interviewees' educational experiences with notebooks. Here we rely on the interviews in a non-research capacity to motivate and reinforce our pedagogy recommendations regarding documentation. For the purposes of this paper, we focused narrowly on the responses to the second section of the interview. The data from the third section have been reported on in our previous work.\cite{Stanley2016a}

Prior to the interviews, the interviewees were asked to find an entry they wrote in their research notebooks that they evaluated as higher quality and one of lower quality, and to bring this notebook to the interview. During the interview, the interviewer went through these entries in detail with the interviewee and had them compare and contrast the content and features of both entries. This was done as a way for us to understand the interviewees' documentation through specific examples, as well as provide a visual aid for discussion about what does and does not make for good scientific documentation. 

To synthesize the interview data, we listened through each interview individually and wrote field notes that summarized the researchers' responses. We then went through all of the field notes to look for aspects of their documentation practice that were either variable or consistent amongst the interviewees. Our recommendations were based on these findings and were informed by our experience as physics researchers and lab instructors. In the following sections, we outline the components that are important features of documentation and describe our recommendations for incorporating these into a lab course. We refer to the interviewees as ``researchers'' for the remainder of the paper.

\section{Purpose of lab notebooks}
\label{sec:purpose}

Researchers generally had a consensus view of the purpose for lab notebooks in their research. The notebook is intended to serve as a record of precisely what one did (both successfully and unsuccessfully) throughout the course of one's experiment---it was described as being the memory of the experiment. It was emphasized that the complexity of their experiments made it too difficult to remember all of the daily details, and so the notebook was essential for keeping track of them. Additionally, in order to make progress in their work, researches had to synthesize together results from different days---by comparing and contrasting different measurements, they were able to make sense of the subtitles of their experiment. This required that the details of various days' efforts were adequately recorded. These different measurements may have been taken over the course of days, weeks, or months. Also, from a summative perspective, information in the notebook was essential to corroborate anything that would ultimately be published. For many of the researchers, the notebook also serves as a place to develop new ideas for the future trajectory of the project, so that they could revisit and further refine these ideas as new results arose. Furthermore, the notebook served to communicate the researcher's efforts to others involved in the project either at present or to those in the future. In essence, the purpose of the notebook was to help make sense of the experiment, think through future directions of the project, create a foundation for publications, and communicate progress with others.

\section{\label{sec:principles}General characteristics of documentation}

In this section, we discuss the general characteristics of how the researchers approached the process of documenting their experiments. In an attempt to synthesize their approach, we focus on the aspects that seemed the most germane to the broadest range of environments. We describe the format and structure of researchers' entries, three guiding principles researchers used when documenting, the different types of information researchers record, and common difficulties researchers face when documenting.

\subsection{Format and structure of entries}

Although researchers generally agreed on the purpose of using a notebook for research, this does not necessarily mean their actual documentation process was similarly consistent. In fact, we found a large degree of variation between researchers in their notebook entries. This variation was in both the specifics of the recorded information, and in the overall structure and format of the entries.

As one might imagine, the specifics of the information in the researchers' notebooks were largely dependent on the specifics of the researchers' experiments---differences in research focus, experimental design, equipment, and types of measurements naturally lead to differences in the information that was recorded. 

Furthermore, no two formats for their entries were the same, from researcher to researcher---the layout of the entries, the length/detail of the entries, how information like graphs/tables/data were arranged, how neat and organized overall were the entries, etc, were all diverse.

However, there were some notable commonalities among the researchers' notebooks. Namely, all notebooks were organized chronologically; the start of each entry was dated; relevant data files were clearly labeled and referenced; and figures/graphs had appropriate units, labels, and legends. But aside from these more basic considerations, there was great diversity in the appearance of the researchers' entries.

\subsection{Principles of context, audience, and timescale}
\label{sec:contaudtime}

Although the structure and format of researchers' notebooks are diverse, there were several general principles for documentation that were implemented by virtually all the researchers. In part, these principles encapsulate the broader picture of the purpose of scientific documentation and how it facilitates research. We labeled these as \emph{context}, \emph{audience}, and \emph{timescale}.

\emph{Context}: Understanding the context means understanding the \emph{what} and the \emph{why} of each experimental decision---that is ``\emph{what} was it that I measured and \emph{why} did I measure it?'' In other words, it means understanding each notebook entry in the broader context of the entire experiment. Researchers considered the inclusion of context to be a characteristic of good notebook entries. Conversely, they considered it bad practice to simply write down the numbers for each parameter and list the different data they collected without explaining their reasoning, since this made it more difficult to interpret their measurements in the future.

\emph{Audience}: The next principle is that of audience. For most researchers, they themselves were the primary audience of their own notebook. However, given that most research is collaborative, the researchers expected that the lab records might be used by others involved in the experiment. This included peers in the same research group, the advisor, researchers from collaborating research groups, or researchers involved with the project in the future. Researchers emphasized that it was important to consider who the audience would be for the information they were recording. In other words, it was important for them to be aware of what they assumed without recording, and considering whether or not others may be able to make sense of the context of the entry without this information. Some of that information might be relevant only to themselves (such as information that would be used only that day), but some information would be referenced by all members of the research group and therefore needed to be broadly understood. Many researchers did not have to consider a broader audience until they began working in an authentic research setting.

\emph{Timescale}: The final principle is that of timescale. Researchers described a broad range of different timescales on which they wrote and referenced their documentation. The recorded information could be referenced in a week, a month, or even more than a year from when it was written down. Researchers describe looking back on notebook entries they had written months or years before, and in some cases looking back on the notebooks of former researchers from up to half a decade before. It was important to consider what the relevant timescale was for any piece of information and writing it down in sufficient detail that reflected that timescale. Some of the information might have been of more short-term importance (\emph{e.g.}, equipment parameters that will be updated in the subsequent few days), whereas other information researchers would continue to come back to over the course of months (\emph{e.g.}, a commonly reproduced alignment procedure). Entries used on shorter timescales generally did not need to be as thorough and detailed as that which would be revisited months later.

In the subsequent sections, we discuss how these principles of \emph{context}, \emph{audience}, and \emph{timescale} played a role in the types of information found in researchers' notebook and how they were implicated in the difficulties of performing documentation.

\subsection{Types of information}
\label{sec:info}

Although the specific information in researchers' notebooks was largely dependent on their experiments, all the information could be categorized in one of five ways. These categories are as follows: 

\emph{Objective information}: This consists of the parameters, settings, and data that result from measurements, alignments, etc. This type of information is likely what undergraduate students imagine to be contained in scientific records. The objective information might be thought of as the “facts” of the experiment---the core of the documentation.

\emph{Analytical information}: Commonly, researchers performed partial or incremental analysis on raw data throughout the entire experimental process. Often this was done to directly compare experimental results to theoretical models/predictions. The information from this analysis is often recorded in the notebook alongside the objective experimental details about the data.

\emph{Interpretive information}: This usually manifests as the researcher's interpretation or evaluation of the objective information. Interpretation of measurements (\emph{e.g.}, ``this data looked unusual'' or ``it seems like the alignment is bad'' or ``I think this measurement is better than the previous one we took.''), helped the researchers recall their in-the-moment impressions of the data when reflecting on past entries (incorporating the principle of \emph{timescale}). Furthermore, it helped to make sense of the measurement relative to other similar measurements (incorporating the principle of \emph{context}). Finally, this interpretation helped other researchers, who did not collect the data, make sense of the measurements (incorporating the principle of \emph{audience}). 

\emph{Synthesis information}: This information served to connect the daily data and documentation to the broader scientific questions of interest, by speculating on how the data may be relevant to the working hypotheses or theories. This information helped the researchers to formulate new hypothesis, which then could inform the direction of future inquiry. This information could be thought of as incremental ``conclusions'' about the results---it is speculation about what the data ``means'' and how it fits into the broader picture of the experiment. This serves to synthesize the previous three types of information and is one of main ways that researchers incorporated the principle of \emph{context} into their documentation.

\emph{Brainstorming information}: This consisted of the researchers' future plans or directions for their experiments. This information entailed both short-term/incremental plans (\emph{e.g.}, taking more data similar to previous measurements, but with slightly different parameters) as well as long-term/substantial plans (\emph{e.g.}, complete redesigns of experimental apparatus). Researchers constantly reflected on and re-conceptualize the day-to-day outcomes of their work. Thus, the researchers provided an overarching narrative for the experiment, which was useful for the primary researcher not only to keep track of ideas, but also to communicate to other researches how the experiment was evolving---consideration for the principles of \emph{context} and \emph{audience}.

These were the types of information found in all the researchers' notebooks. However, the balance between these different types was variable from one notebook to the next. Some researchers kept brainstorming and analysis to a minimum in their notebooks and instead predominantly wrote objective information, where as others provided detailed narratives of their real time thought process that included a great deal of synthesizing and future planning.

\subsection{Difficulties with documentation}
\label{sec:difficulties}

Researchers described some common difficulties that they experienced while developing their documentation skills. These difficulties were present at all levels of experience, and many were persistent struggles that they still confront on a daily basis. These difficulties can be seen as relating to attempts to incorporate the principles of \emph{context}, \emph{audience}, and \emph{timescale}.

\emph{Time investment}: The process of keeping lab records is a fine balance between writing enough detail so that the records will be useful in the future and doing so in a time efficient way that does not slow the progress of the experiment. Given that record keeping is a time intensive part of the experimental process, many researchers expressed feeling they should be taking more time to write additional information in the records they keep. Very few felt they spend too much time adding detail to these records. Many expressed difficulties in finding enough time to write down all that they felt they should.

To adaquately incorporate the context of the daily activities so that the entries are understood by the intended audience, the reseracher includes more detail in their entries. Fundamentally, this time investment is a resource challenge that will persist throughout the experimental process.

\emph{Ambiguity in what to record}: Even when time is not a factor, it may not be immediately clear what is and is not important to record. Most of the researchers expressed that frequently, when writing information down, they were not necessarily sure if that information would be useful in the future. If time was pressing, deciding what to record could be even more of a challenge. Several researchers described examples of instances where they later realized that not writing something down caused them to have to retake measurements, costing them a great deal of time and effort. At the time, the researchers were uncertain about the importance of this information they chose not to record.

Generally, the researchers felt it better to err on the side of more information, not less. As they gained more experience/familiarity with their experiment, they came to understand what information was and was not needed. In part, this ambiguity is the researcher's lack of understanding of the context of the experiment, which prevents them from incorporating that context into their entries.

\emph{Delayed Gratification}: The entries of most notebooks are rarely intended to be immediately beneficial. Consistent with the stated purpose of the notebook, the entries are predominantly beneficial in hindsight, after a good deal of complimentary information has been collected. For this reason, researchers described that it could be difficult in the moment to be diligent in recording all the pertinent information, especially given the time consuming nature of documentation.

This difficulty relates to the consideration for timescale and audience. It can be difficult for researchers to find the motivation to document adequately when the information they record is not intended to be useful for them directly, not going to be utilized until far in the future, or both.

\emph{Cause of bad entries}: Poor entries commonly occurred at points when the researchers were not sure about the state of their experiment (\emph{e.g.}, how to interpret the data they were getting, how they should proceed, whether or not they were having technical issues), either because they were getting confusing results or they were involved in new experimental activities with which they were not familiar. Because the experiment was not proceeding as expected, they either did not know what to write down or felt it was a waste to write down things they did not know how to interpret. However, given the basic purpose of the notebook, good documentation during these points in the experiment might eventually be fruitful upon reflection at some point in the future. What might not make sense in the moment, may later make sense when more information has been collected and the context is better understood.

\section{Recommendations for instruction}
\label{sec:recommendations}

\begin{figure*}[t]
\centering 
\includegraphics[scale=0.85]{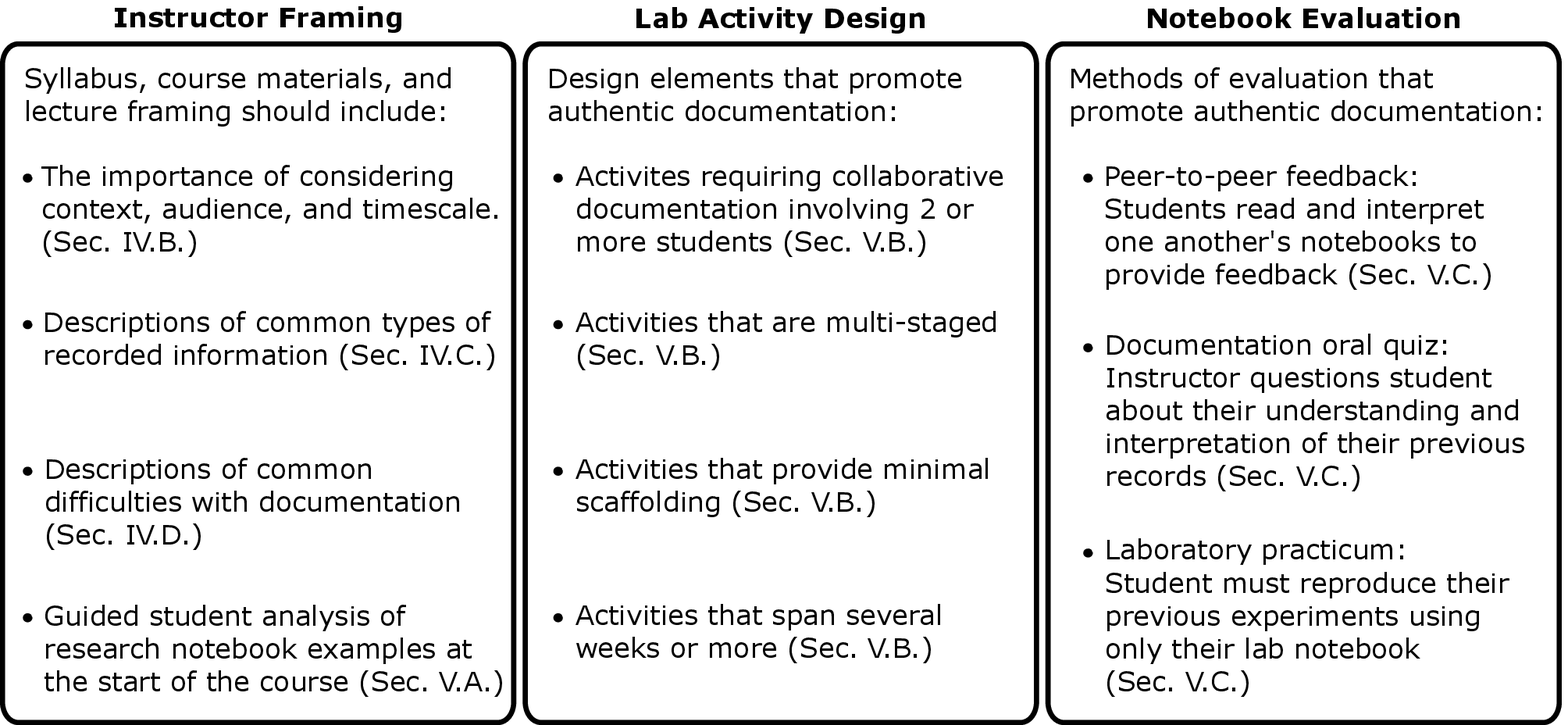}
\caption{Notebook recommendations. Here we present a summary of our recommendations for incorporating authentic documentation into a lab course. Not all of these will be appropriate for all courses, so we encourage instructors to adopt those they feel will be best suited to their learning goals and course design. For each point, we have included reference to the relevant section containing further detail.}
\label{fig:recommendations}
\end{figure*}

In this section, we discuss the recommendations for instructors that incorporate the features of researchers' documentation practice, discussed in Sec. \ref{sec:purpose} and \ref{sec:principles}. These recommendations consist of three parts: instructor framing, lab activity design, and evaluation/grading of notebooks. Instructor framing consists of the ways in which the instructor outlines and presents the use of documentation in their course. Lab activity design addresses design elements that could be incorporated into students' lab activities, which may be part of, or ancillary to, the lab experiments. Evaluation consists of recommendations for how instructors could evaluate and grade student notebooks in a way that effectively encourages productive use of notebooks. A summary of these recommendations can been seen in Fig. \ref{fig:recommendations}. At this point, we would like to acknowledge the great degree of variation in course design, equipment availability, class size, institution type, learning goals, etc., from lab course to lab course, that might make it difficult to adopt very specific recommendations. For this reason, our recommendations herein instead consist of broad ideas intended to be adapted for any particular lab course context.

\subsection{\label{sec:framing}Instructor framing}

Instructor framing is a valuable first step to teach students the importance of, and approach to, scientific documentation. There are a number of ways that the features of researchers' documentation could be presented to students. For example, they could be organized into a notebook primer and disseminated to students at the start of the course. Alternatively, they could be the basis for instructor-guided student discussions about the role of documentation in lab. The following are suggestions to communicate to students when framing how students should make use of documentation in their lab course.

\emph{Purpose}: Having a clear idea of the role that the notebook plays in lab can help to motivate students to engage in and develop their documentation skills. As described in Sec. \ref{sec:purpose}, the notebook serves a number of purposes in lab and students should be made aware of all of these, so that they do not limit their expectation of what constitutes authentic scientific documentation. Providing students with concrete examples for each purpose could help them to better conceptualize how they might approach their own documentation.

\emph{Format and structure}: It is not consistent with authentic practice to require a rigid format for student notebook content or structure. To the contrary, instructors should expect that the content and style of what students record in their notebooks will vary from activity to activity and lab course setting to lab course setting. Thus, it is likely counter-productive to the development of authentic documentation skills to require students to conform to a specific format or template for their notebook entries, aside from the few basic structural similarities stated above. Moreover, within this basic structure for notebook entries, students should be encouraged to adapt and develop their own documentation format that best suits the requirements of their lab activities---practice that will aid them in future research settings. However, if a choice is made to standardize elements of the format (\emph{e.g.}, for grading purposes), the purpose of the standardization should be communicated to the students. To the extent that a format is imposed, it will benefit the student to vary the format from activity to activity in order to provide opportunities for students to practice different styles of documentation suitable to different experimental contexts.

\emph{Context, Audience, Timescale}: Instructors should explicitly highlight these three principles (outlined in Sec. \ref{sec:contaudtime}) and prompt students to consider them when performing documentation during their lab activities. First, for effective documentation, students must understand and make explicit the context of the information they are recording. So, when recording information in their notebooks, students should be encouraged to consider whether or not they can understand how the documented details pertains to the experiment as a whole. If they are simply writing down the numbers for their parameters and data, without explaining the reasoning behind the measurements, they may be unable to make sense of what is written. This requires that the students are actively engaged in understanding the motivation for their measurements and interpreting the results they obtain.

Second, in order to encourage students to be mindful of the audience for their notebook, they should imagine how their writing may be interpreted by others---being aware of the things they infer or assume, without writing it down, and ask whether or not others may be able to make sense of the context of the entry without this information. In a lab class, the audience will primarily be the students themselves but may be their lab partners and the instructors as well. These different audiences are likely to have varying degrees of familiarity with the experiment, so will need varying degrees of recorded detail to make sense of the documentation.

Third, it should be conveyed to students that in authentic scientific documentation, notebooks are written in and referenced on a wide range of different timescales. This range depends on the nature of the information being recorded and the stage of the experiment. They should be encouraged to ask themselves, ``When might I need this?'' whenever writing down new information. The more time that elapses between when one writes the information and when it is referenced, the more detail they will need to include in order to understand the context of the notebook entry.

\emph{Types of information}: Though students might find obvious some of the categories in Sec. \ref{sec:info}, many will not realize the remaining categories are not only appropriate, but also beneficial to document. Students may naively assume that only objective and analytic information is suitable for scientific documentation and might not recognize the benefit of providing some of their own interpretation, synthesis, or future planning in their entries. The relevance of these other types of information could be easily outlined for students in a primer, early in their course. These lend themselves to instruction even for structured introductory labs. To improve students documentation, it is likely helpful to describe to them these categories of information, explain their function in the notebook, and outline how they are related to the principles of \emph{context}, \emph{audience}, and \emph{timescale}. In this way, students are provided a framework for how to think about the range of information they might record, but still have the opportunity to think critically about the specifics of what they should write.

\emph{Difficulties}: Instructors should communicate to the students the common difficulties that researchers experience, as described in Sec \ref{sec:difficulties}. Students should be encouraged to not look at record keeping as an afterthought to the actual experimental process, but rather as an integral part that will require substantial amounts of lab time to get right. It should also be communicated to them that they will not necessarily know what is and is not important to document and that most of the benefit of the information they record will be realized only at some point in the future, upon reflection. If students are made aware that these difficulties exist and are common to researchers' lab experiences, it may help them to better navigate the frustration and discouragement they might experience while attempting to keep suitable lab records. In turn, this may help improve their documentation, and thereby aid them in completing their lab activities. Ultimately, this could result in a better overall lab course experience.

The instructor's framing could be reinforced by providing the class with examples of actual scientific notebooks (\emph{e.g.}, notebooks from the labs of physics faculty in the department) and having the students analyze them as a lab activity. The students could be asked to describe things like the types of information they see, how information is organized, etc, in order to get them thinking about documentation more broadly. By using multiple examples of notebooks, students could compare and contrast the details between them. A similar approach to framing documentation has been described in other work.\cite{Atkins2013,Salter2013}

\subsection{Design of lab activities}

Although thoughtful framing of lab notebooks, in the ways outlined above, can go a long way to providing a foundation for students to understand what constitutes scientific documentation, by itself, it is unlikely to cause students to engage in, and develop, productive documentation practices of their own. Essentially all the researchers emphasized the importance of having authentic use of notebooks for documentation in lab courses for the students to effectively learn scientific documentation skills. Specifically, the design of the lab activities should explicitly incorporate student documentation. If lab activities are not designed in such a way that necessitates use of lab notebooks, commensurate with how scientific documentation is performed, then students are not likely to develop this skill---regardless of how well the theoretical use of lab notebooks is framed at the start of the course. In previous work, we discussed that many of the interviewees felt keeping a lab notebook was not practically useful for the activities in their undergraduate lab courses and as a result they did not improve their practice as much as they otherwise could have.\cite{Stanley2016a} An example of this is expressed in the following quote.
\begin{itemize}
\item[]{\emph{``The usefulness of a lab notebook and the utility of it only became apparent to me when I started doing long term research that spanned days, weeks, and months and was not canned. Only then did I really start relying on the notebook to make sense of all the unknowns. Those unknowns weren't inherent to the activities I was doing in [my lab class], so it didn't really make sense to have a lab notebook in the way that I do now. But to fix that, or to make my experience more authentic in the way that we just described, it requires more than just grading the lab notebook better or having lab notebook requirements be better. It would involve fundamentally changing the structure of the experimental course, as a whole.''}}
\end{itemize}

In order to design lab activities that provide productive learning opportunities for documentation, the lab instructor should incorporate consideration for authentic \emph{context}, \emph{audience}, and \emph{timescale}. To incorporate these aspects, lab activities could: 
\begin{enumerate}
\item{involve multiple students in such a way that they must rely on one another's documented work to make progress on the broader goal of the lab activity.}
\item{be sufficiently complex and multi-staged so that students must make sense of, and synthesize together, a number of different measurements.}
\item{provide the minimum amount of scaffolding in the lab guides so that students must be proactive in recording the step-by-step process of the activities themselves, instead of relying on a lab guide.}
\item{span a sufficiently long period of time so that students rely on more than their memories or hastily written notes to make sense of the results.}
\end{enumerate}
Activities that address one or more of these features will provide students with an experience that emulates facets of authentic documentation.

One example would be a project-based lab course where students work in teams of 2--4 on a single project for much or most of the academic term. These projects could be chosen by the students (the oversight of the instructor could assure that the project is sufficiently complex but tractable). Open-ended projects that lack thorough lab guides to rely on, provide students the opportunity to experience what it is like to rely on their own documentation. Furthermore, if the project has a number of different components, this provides the students the opportunity to divide up this work and practice documenting the progress of their respective efforts, which can then later be shared with the rest of the group when the different components of the project are to be synthesized. In this way, students get experience reading other's documentation and could be encouraged to provide feedback about the clarity and thoroughness of the records. This type of interaction can authentically mimic how documentation is performed in actual research settings.

Alternatively, a more structured approach may be to have a project-based lab course in which each group of students spends some period of time (a few weeks, say) working on each project. The goals of these projects could be outlined by the instructor. Every project would require the entire academic term to complete, but each group would be completing only a portion of each one. When the groups rotate to a new project, they must rely on the lab records, documented by all previous groups, in order to understand the progress that had been made on the project and how they should proceed. In this way, the efforts of the previous groups get conveyed through the project's notebook and the current group must be mindful of the documentation they contribute, in consideration for the groups that follow them. A similar activity has been suggested in the literature.\cite{Graetzer1972}

However, there are also activities that can be incorporated into very structured lab courses that provide exposure to aspects of authentic documentation. One example would be to have lab activities that require students to compare their results to those of previous students in the class. The students would identify and assess the previous results by looking at the previous students' notebooks. The relevant portions of the notebooks would be identified and provided by the instructor. These activities could consist of the calibration/alignment of some apparatus or the completion of a multi-stage measurement, for example. Designing these activities so that students would have to examine the parameters and conditions under which the previous students collected their data, would encourage a greater engagement with the documentation than merely comparing a single number. This exercise would be helpful in analyzing the context of entries and providing the students an authentic experience with extracting information from documentation. This could prime them to consider how they go about record information in their own notebooks.

It should be noted that these suggestions for structured lab courses provide students with experiences that get at the principles of context and audience, but not timescale. The short-term nature (the duration of each activity is at most a couple weeks) will likely make it difficult to address the principle of timescale in such an instructional environment.

In this section, we have presented a number of general ideas for designing lab activities that actively promote aspects of authentic documentation practice. The design ideas presented in this section are far from the only ways in which documentation can be addressed in lab courses. With some creativity and mindfulness about authentic use of notebooks, instructors can come up with ideas that best align with their course---when doing so we recommend that instructors keep in mind the four features that we have enumerated above in this section.

\subsection{Evaluating student notebooks}

The last component is how to grade/evaluate students' notebooks. When asked how to evaluate students use of lab notebooks, the researchers responses were quite universal. Both what they did and did not suggest was informative. None of the interviewees stated that the student should be evaluated on the format, neatness, organization, or degree of detail of the entries. Instead, the focus was on holistic and pragmatic aspects of the notebook. Specifically, the researchers stated that students' entries should be interpretable as a whole and make sense in the broader context of the notebook. They stated that the student should be able to use their notebook entry as a guide to explain what they had done that day to someone else. If a student had good documentation, they would be able to open their notebook to an arbitrary page, explain what occurred on that day, and provide rational for why those actions were germane to the experiment as a whole. 

One way of accomplishing this would be peer-to-peer analysis of notebooks. By pairing up students and having them read through one another's notebook, the students would attempt to make sense of the experiment using only the documented information and provide feedback about its understandability.  The instructor could facilitate this activity by providing questions the students could take into consideration when analyzing another's notebook (\emph{e.g.}, What was the intended goal of their experiment? What measurements did they make? Do they provide interpretation for the results of their measurement? What difficulties and uncertainties were they confronted with during their experiment and what did they do to proceed?). Ultimately, this would be beneficial to both students: the student receiving feedback would get direct input about how their documentation is being interpreted and the student providing the feedback would get experience in what it is like trying to make sense of scientific documentation.

Another approach would be for the instructor to give oral examinations in which they question students about their understanding and interpretation of specific notebook entries. The questions asked would be similar to those described in the previous paragraph, in order to see how well students can make sense of their own documentation. This could be a midterm or end of term examination, which would allow the instructor to select notebook entries from a range of timescales. A similar idea is suggested,\cite{Gould1927} except on a much shorter timescale, which does not engage students in examining their records on timescales common to research (weeks and months).

Finally, students could be instructed to perform a lab practical, in which they must reproduce an experiment they performed and documented earlier in the course, using only their lab notebook as a guide.  In essence, this would get at one of the fundamental purposes of documentation---reproducibility.

Though these approaches might be time consuming, it would align the evaluation/grading of the notebook with authentic documentation practice,  communicate to students why documentation is important, and  encourage them to actively reflect on how they record their entries.

\section{In summary}
\label{sec:sum}

Informed by interviews with graduate physics researchers, we have presented recommendations for which aspects of authentic documentation are important for students and what to consider when incorporating the use of lab notebooks into a lab course. The general principles that can guide good documentation are (1) to convey \emph{context} in one's entries, (2) to be aware of the \emph{audience} of the entry, and (3) to take in to consideration \emph{timescale} of use for recorded information. These three principles help to inform the types of information to be recorded, navigate potential difficulties with documentation, and orient students' view of the purpose of maintaining a lab notebook. Finally, the design of lab activities is important for students to effectively learn how to perform scientific documentation. Lab activities must be designed so that students rely on their notebooks in a manner similar to an authentic research setting. Regardless of how well instructors frame the use of notebooks, or convey their importance to experimental physics, if the course design does not make authentic documentation an integral part of the lab activities, it is unlikely students will make significant progress toward developing effective documentation skills.

\begin{acknowledgments}
We would like to acknowledge and thank Robert Hobbes, Dimitri Dounas-Frazer, and Bethany Wilcox for their helpful feedback. This work was supported by NSF grant nos. DUE-1323101 and PHY-1125844.
\end{acknowledgments}
%
\bibliographystyle{unsrt}



\end{document}